\documentclass[aip,reprint]{revtex4-1}

\usepackage{amsmath}
\usepackage{graphicx}

\begin{document}

\title{
Two famous results of Einstein derived from the Jarzynski equality}

\author{Fred Gittes}

\affiliation{Department of Physics and Astronomy, Washington State University,
Pullman, WA 99164-2814}

\date{\today}

\begin{abstract}
The Jarzynski equality (JE) is a remarkable statement relating
transient irreversible processes to infinite-time free energy
differences.  Although twenty years old, the JE remains unfamiliar
to many; nevertheless it is a robust and powerful law.  We examine
two of Einstein's most simple and well-known discoveries, one
classical and one quantum, and show how each of these follows from
the JE.  Our first example is Einstein's relation between the drag
and diffusion coefficients of a particle in Brownian motion.  In
this context we encounter a paradox in the macroscopic limit of the
JE which is fascinating, but also warns us against using the JE too
freely outside of the microscopic domain.  Our second example is
the equality of Einstein's $B$ coefficients for absorption and
stimulated emission of quanta.  Here resonant light does irreversible
work on a sample, and the argument differs from Einstein's equilibrium
reasoning using the Planck black-body spectrum.  We round out our
examples with a brief derivation and discussion of Jarzynski's
remarkable equality.
\end{abstract}

\maketitle

\newcommand{\KB}{{ k_{\text{\tiny B}} }}
\newcommand{\KBT}{{ k_{\text{\tiny B}} T }}
\newcommand{\avg}[1]{{\langle {#1} \rangle}}
\newcommand{\sssdnarrow}{{\scriptscriptstyle\downarrow}}
\newcommand{\sssuparrow}{{\scriptscriptstyle\uparrow}}

\section{Introduction}
\label{Introduction}

Suppose a system initially in equilibrium at a temperature $T$ is
transiently disturbed, perhaps dramatically, over some limited
interval of time.  The disturbance will incur some amount of external
work $W$, and it may leave the system in some new state.

Since the microscopic state of the system is a matter of probability,
the irreversible work done in any particular trial will vary.
Furthermore $W$ certainly depends on details of the disturbance,
and some or all of it may be dissipated as heat.  For these reasons
one would expect $W$ to be only loosely connected with the equilibrium
eventually reached.  The Jarzynski equality (JE) confounds this
expectation with the exact average
\cite{Jarz1997a,Jarz1997b}
\begin{equation}
    \avg{e^{-\beta W}}
    \;=\;
    e^{-\beta\Delta F}
\label{911}
\end{equation}
where $\beta = 1/\KBT$
and $\Delta F$
is the change in Helmholtz free energy at an infinite time, after
the system has found its new equilibrium.  It is remarkable that
$\Delta F$ is so directly linked to the history of the disturbance.
The Jarzynski equality holds both classically and quantum-mechanically,
and generalizes to cases where the initial and final temperatures
are not equal (Section \ref{Derivation}).

Eq~(\ref{911}) is a remarkable statement that invites explanation
and application.  Over the past two decades the JE, along with its
generalization to the ``fluctuation theorems'' of Crooks,
\cite{Crooks1998,Crooks1999,Crooks2000} Evans and Searles
\cite{EvansSearles2002} and others, have been important in such
areas as calculational and experimental molecular dynamics.\cite{Hummer2001,
Bustamante2005, Harris2007,Jarz2011}  On the pedagogical side,
however, the JE remains exotic, and absent even from otherwise
excellent textbooks on statistical mechanics.\cite{Peliti2011}
Helpful expositions of (\ref{911}) have included a very nice
application to gas in a piston,\cite{LuaGrosberg2005} and interesting
applications to systems of springs\cite{Hijar2010} and to magnetic
resonance phenomena.\cite{Ribeiro2016}  These and other treatments
have been valuable and rewarding to work through.  Nonetheless,
illustrations of the JE have sometimes struck the author as somewhat
artificial, or tangential to the mainstream of physics.

Seeking evidence for the centrality and robustness of the JE and
fluctuation theorems in areas of physics he currently teaches, the
author examined two of Einstein's widely known discoveries, one
classical and one quantum.  Both results, while simple, are of deep
significance.  To the author's surprise, it became evident that the
Jarzynski equality implies each of these results.

The first example to be considered is Einstein's relation between
diffusion and drag coefficients, central to Brownian motion and
historically the first fluctuation-dissipation theorem, or
FDT\cite{Landau,Reif,Marconi2008} (different from the fluctuation
theorems mentioned above).  A connection between FDTs and the
Jarzynski equality was previously noted by 
L.~Y.~Chen,\cite{Chen2008a,Crooks2009} but our diffusion-drag relation
is especially useful not only for its simplicity, but because it
illustrates a paradox (Section \ref{Macro limit}) of a type noted
by Jarzynski and others,\cite{JarzRare2006,LuaGrosberg2005} in
the macroscopic limit of the average in Eq~(\ref{911}). Such paradoxes
are an important warning against using the JE as freely as one might
use, say, the Gibbs distribution.

Our second example is Einstein's conclusion of the necessity of
stimulated emission, and the equality of the coefficients of
absorption and stimulated emission.  Einstein's original argument
is based on Planck's black-body spectrum,\cite{Eisberg} and is a
staple of undergraduate courses on modern physics.  Stimulated
emission is also derived in advanced quantum mechanics as a consequence
of the quantum statistics of boson fields such as photons.\cite{ZimanAQM}
We obtain the result from the Jarzynski equality by viewing absorbed
and emitted quanta as work done on a thermal system of two-state
atoms, whose free energy change is straightforwardly calculated.
A surprising feature is that this argument does not explicitly invoke
the Bose nature of the quanta.

Derivations of the Jarzynski equality in the literature vary
in length and accessibility. In Section \ref{Derivation} we provide
an adaptation of an efficient quantum-mechanical derivation of the JE and a
generalization of it, both due to D.~N.~Page.\cite{Page2012}  Finally,
in Section \ref{Derivation} we mention the importance of time
reversal as the underlying basis of the Jarzynski equality.

\section{Strategy for applying the JE}
\label{Strategy}

In each of our examples, our strategy will be to rewrite the average
$\avg{e^{-\beta W}}$ on the left side of the JE by using the leading
terms of a cumulant expansion, a statistical device useful throughout
theoretical physics.\cite{Peliti2011}  In our case this expansion
is
\begin{align}
    \avg{ e^{-\beta W} }
    &\;=\;
    e^{-\beta \overline W
    +(\beta^2/2)(\overline{W^2}-\overline W^2) + \dots }
    .
\label{067}
\end{align}
The quantity $\overline{W^2}-\overline W^2  = \overline{\Delta W^2}$
in the exponential is the variance, or mean-squared variation of
$W$.  Eq~(\ref{067}) is related to the statistical statement $\avg{e^{X}}
\ge e^{\avg X}$, which when applied to the JE to yields the
thermodynamic law $\Delta F \le W$ as a consequence of the Jarzynski
equality.\cite{Jarz1997a}

Eq~(\ref{067}) can be verified if desired, by expanding $e^{-\beta W}$
in a series, averaging the terms, taking a logarithm and then
expanding this again in series.

We make a few additional points.  Derivations of the JE like that
of Section \ref{Derivation} assume a specified time-dependence of
the system Hamiltonian $H$, quantum or classical.  This is not a
strong restriction, however.  For example, any classical force
$f(t)$ is representable in $H$ by a term $-xf(t)$.  Such a classical
$f(t)$ is used in our first example below.  In the derivation of
Section \ref{Derivation} we see how work is defined quantum
mechanically to arrive at the JE.  Our second example below will
introduce quantum work consistent with that definition.

The use of a specific temperature in Eq~(\ref{911}) implies contact
with a heat bath; one might infer that the JE is a feature of open
systems.  In fact, the quantum JE derivation in Section \ref{Derivation}
assumes and requires a closed system.  Thus, strictly speaking,
Eq~(\ref{911}) is being applied to the system of interest together
with its heat reservoir.  For a small closed system without a
reservoir, irreversible work done on a closed system may in fact
leave it at a changed temperature, and this situation is handled
by the generalized Jarzynski equality in Section \ref{Derivation}.

\section{Einstein's diffusion relations from Jarzynski's equality}
\label{Diffusion}

Our first and simplest application of  the Jarzynski equality is
to a particle executing Brownian motion in a viscous medium, a
problem considered by Einstein in 1905.  We seek to establish
Einstein's relation relating  the drag coefficient of the particle
and its diffusion constant, both to be defined below.

To apply the JE, we must construct some protocol in which a variable
amount of irreversible work will be done on the system. We choose
to apply a constant force $f$ to the diffusing particle for a fixed
time $\Delta t$.  In each episode or trial the displacement  $\Delta x$
of the particle, and thus the work done,
\begin{gather}
    W
    =
    f\Delta x
    ,
\label{933}
\end{gather}
is variable.  In the JE we average $\avg{ e^{-\beta W} }$ over
episodes.

\begin{figure}
\includegraphics[width=0.8\columnwidth]
{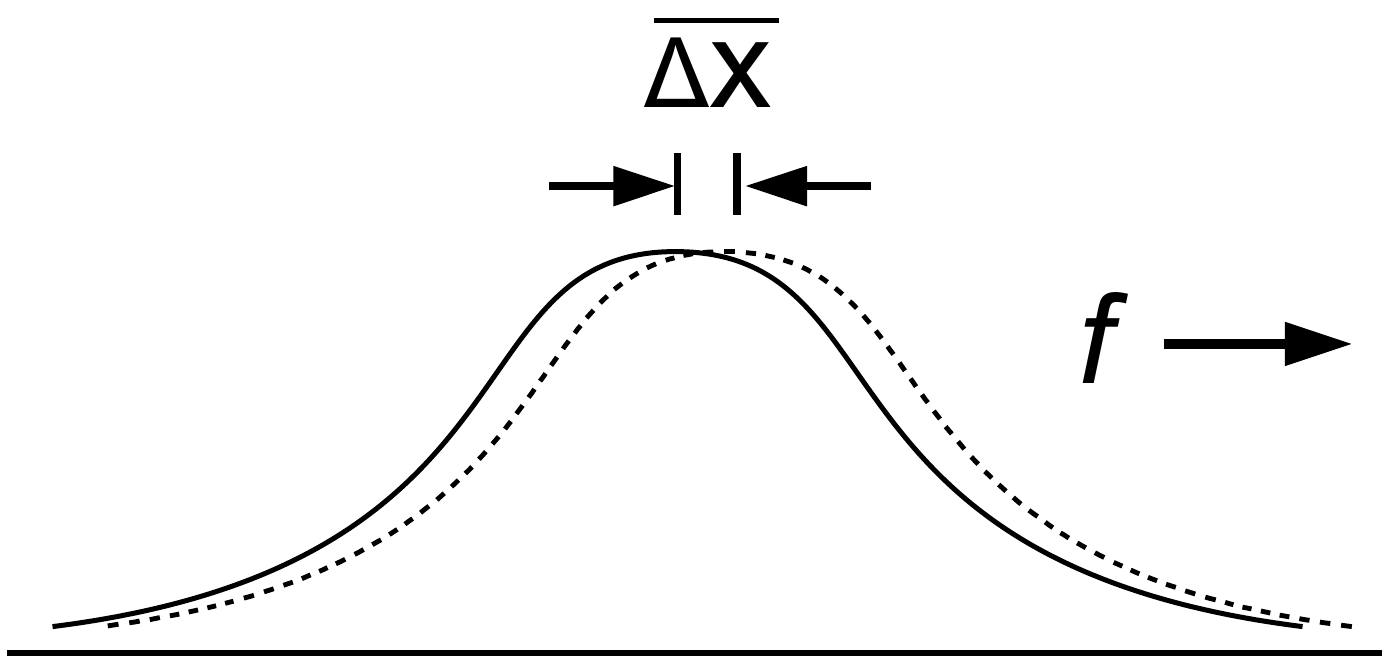}
\caption{
If a small enough force $f$ is applied to a diffusing Brownian
particle, the forced displacement $\overline{\Delta x}$
is much smaller than the width 
$(\Delta x)_\text{rms}$
of the diffusional profile,
so that
$\overline{\Delta x^2}$
can be assumed to be that of free diffusion.
}
\end{figure}

With only dissipative work being done there is no free energy change:
$\Delta F=0$.  Setting $e^{-\beta\Delta F} = 1$ in Eq~(\ref{911}) gives
us the Jarzynski equality for forcing against pure drag or friction,
\begin{align}
    \avg{ e^{-\beta W} }
    &\;=\;
    1
    .
\label{737}
\end{align}
To apply a cumulant expansion we need $\overline{\Delta W^2}$.  In
the limit of small $f$, the forced displacement will be much smaller
than the r.m.s.~diffusion (see Figure 1), so we can assume, to
zeroth order, a free diffusion.  The mean squared displacement of
a free diffusion is $\overline{\Delta x^2} = 2D\Delta t$, defining
the diffusion constant $D$.  Thus
\begin{gather}
    \overline{\Delta W^2}
    \;=\;
    f^2\overline{\Delta x^2}
    \;\approx\;
    2D f^2 \Delta t
    \,.
\label{734}
\end{gather}
Now we can apply the cumulant expansion of Eq~(\ref{067}).  
To lowest order in the force $f$ (and thus in $W$ by Eq~(\ref{933}))
\cite{Reviewersuggest} 
\begin{gather}
    e^{
    -\beta\overline W 
    +(\beta^2/2)
    \overline{\Delta W^2}
    }
    \;\approx\;
    1
    ,
\label{484}
  \\
    \overline{\Delta W^2}
    \;\approx\;
    2\KBT\,
    \overline W 
    .
\label{485}
\end{gather}
Inserting (\ref{734}) and $\overline W = f\overline{\Delta x}$ from
(\ref{933}), we have
\begin{gather}
    2Df^2\Delta t
    \;=\;
    2\KBT\,
    f\overline{\Delta x}
    .
\label{023}
\end{gather}
The drag coefficient $\gamma$ is defined by writing the drag force,
equal and opposite to our imposed force, as $f = \gamma v = \gamma
\overline{\Delta x}/\Delta t$.  Using this in (\ref{023}), we obtain
\begin{gather}
    \gamma D
    \;=\;
    \KBT
    .
\label{720}
\end{gather}
This is Einstein's famous relation between drag and diffusion
coefficients, i.e.~between dissipation and fluctuation.  It is
central to a variety of applications.\cite{Reif,Nelson}  Such
fluctuation-dissipation theorems or FDTs, which we have found to
be a consequence of Jarzynski's equality, arise not only in particle
diffusion but in other phenomena, such as Johnson noise.  A treatment
of voltage, resistance and current would in that case be essentially
identical to what has been done here.  A generic identification of
first-order expansions of the JE with FDTs (essentially our Eq
(\ref{485})) has been noted previously by 
L.~Y.~Chen.\cite{Chen2008a,Crooks2009}
We will not here construct arguments for other FDTs.

\section{Macroscopic limit of the JE}
\label{Macro limit}

Even if the above reasoning is relatively straightforward, the
Jarzynski equality in Eq~(\ref{737}) for a pure drag or friction invites
a fundamental objection.\cite{JTDobjection}  Pulling a macroscopic
object through (say) thick molasses requires a fairly definite
amount of work $W$.  In repeated trials, $e^{-\beta W}$ ought then
to remain close to a value smaller than $1$ in essentially every
trial.  On the other hand, the JE in Eq~(\ref{737}) strictly requires
$\avg{ e^{-\beta W} } = 1$.  How can this be possible?

The resolution of this paradox highlights peculiarities of the
average $\avg{ e^{-\beta W} }$.  In our case of friction or drag,
suppose we drag a block across a surface against friction.  Molecular
motions in the substrate could in principle impart (rather than
remove) macroscopic momentum, reversing the force and making $W$
negative---we would have to restrain the block!  These negative-$W$
trajectories must exist, as time reversals of trajectories we do
observe, but they are suppressed by fantastically small Boltzmann
probabilities of order $e^{-\beta |W|}$, that reflect the entropic
cost of removing energy from microscopic degrees of freedom.  On
the other hand, they contribute enormous values $e^{+\beta |W|}$
to the Jarzynski average that compensate for small probabilities.
In this way, a macroscopic limit of the average $\avg{ e^{-\beta W} }$ 
will be held to its required value by the contributions of
fantastically unlikely scenarios.

Jarzynski has emphasized\cite{JarzRare2006} this dominance of ``rare
realizations'' in $\avg{ e^{-\beta W} }$ and the role of time-reversed
trajectories, as well as the puzzle of the macroscopic limit for
this average in an adiabatic process.\cite{JarzPers2007,CrooksJarz2007}

\section{Stimulated emission from Jarzynski's equality}
\label{Stimulated emission}

Our second application of the Jarzynski equality is to the phenomenon
of stimulated emission.  We want to show that any light-absorbing
two-state system must also undergo stimulated emission, and that
the absorption and emission rates (to be defined) must be equal.

When Einstein considered this problem in 1916, he assumed the light
to be in thermal equilibrium (a Planck black-body distribution).
To apply the JE, we instead do irreversible work on a ``sample''
of $N$ two-state subsystems at temperature $T$ (see Figure 2), by
applying light to it for a fixed time.  The incident light is not
thermal; instead its energy $\hbar\omega$ matches the excitation
energy of the subsystems.

We will write the work $W$ in terms of the number of absorptions
and emissions that occur, in each trial, and as usual evaluate
$\avg{ e^{-\beta W} }$ with a cumulant expansion, to lowest order
in $\beta$ (a high-temperature limit).  Next, in $\Delta F$, we
will need the entropy change $\Delta S$ of the 2-state systems given
the average number of up and down state changes. Finally we will
solve the JE in our high-temperature limit for the stimulated
emission.

\begin{figure}
\includegraphics[width=0.8\columnwidth]
{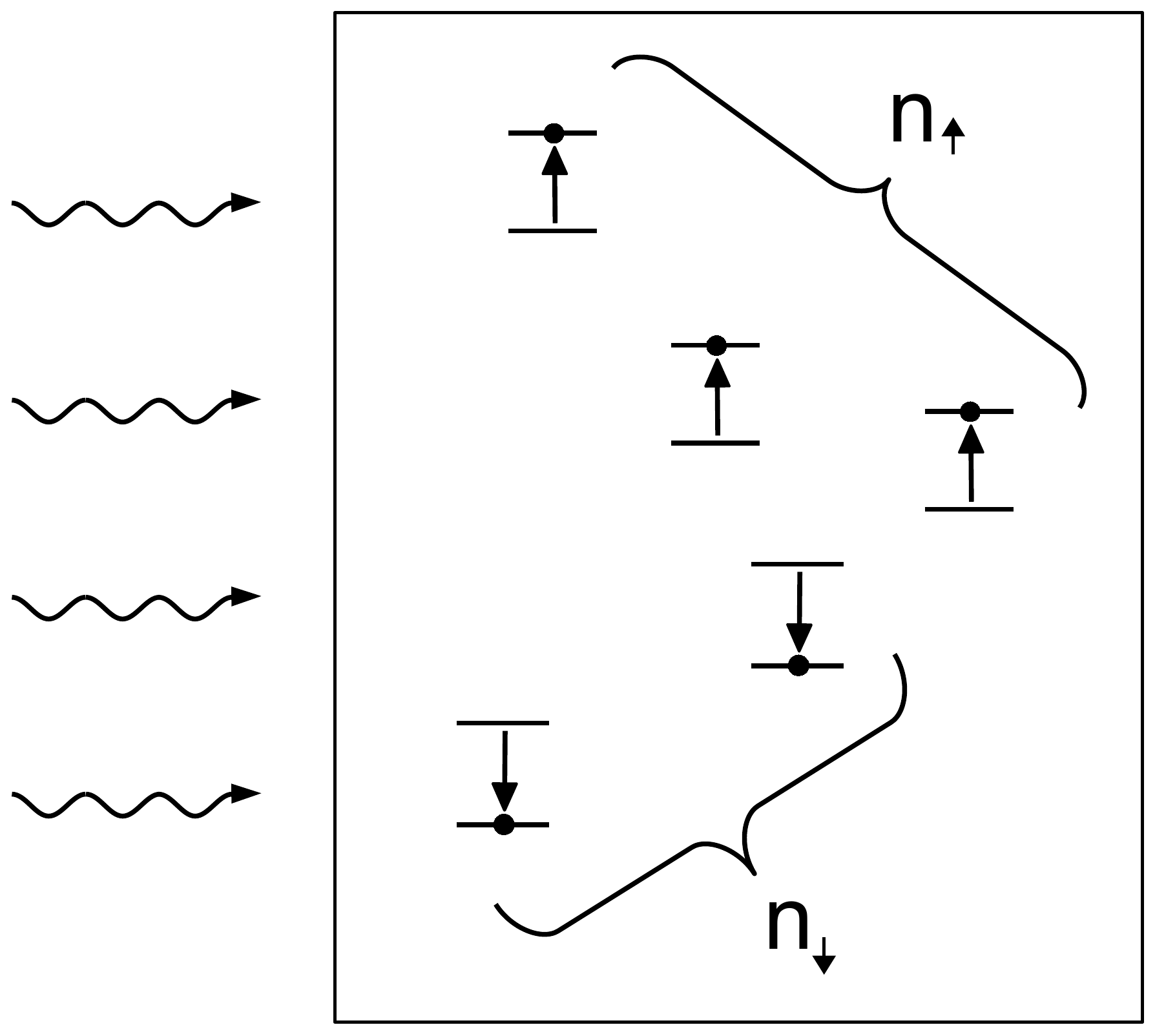}
\caption{
Model system of resonant light with intensity $I$ and fixed duration,
illuminating a collection of thermalized two-state subsystems.  The
subsystems are spatially separated so that their interactions with
the light occur independently.  During each trial, $n_\sssuparrow$
absorptions and $n_\sssdnarrow$ induced emissions occur.
Spontaneous emission is  excluded from 
consideration
by assuming the
thermal re-equilibration of states to be fast.
}
\end{figure}

Initially ${{n}}$ of our subsystems are in the upper state and
$N-{{n}}$ are in the lower state, a Boltzmann distribution.  Under
a light of intensity $I$, the work due to $n_\sssuparrow$ absorptions
and $n_\sssdnarrow$ emissions is
\begin{gather}
    W
    \;=\;
    \hbar\omega
    (
    {n}_\sssuparrow
    {-}
    {n}_\sssdnarrow 
    )
    .
\label{506}
\end{gather}
For our cumulant expansion we will need the variance 
$\overline{\Delta W^2} = \overline{W^2} - {\overline W}^2$.  
Since $n_\sssuparrow$ and $n_\sssdnarrow$ are independent, 
their variances simply add.  Also, as random counts, they are Poisson
variables\cite{Reif} for which the variance is equal to the mean.
Using these facts we can obtain from
(\ref{506})
\begin{align}
    \overline{\Delta W^2}
    \;=\;
    (\hbar\omega)^2
    (
    \overline{n}_\sssuparrow
    +
    \overline{n}_\sssdnarrow 
    )
    .
\label{208}
\end{align}
Now we invoke the Jarzynski equality, Eq~(\ref{911}).  As in our 
previous example, we apply the cumulant expansion of Eq~(\ref{067}) 
to $\avg{ e^{-\beta W} }$.  On the right side of the JE we need the
free energy change of the system, $\Delta F = \overline W - T\Delta S$,
with $\Delta S$ the entropy change due to changes in upper- and
lower-state occupation.  The JE becomes
\begin{gather}
    e^{-\beta \overline W
    +(\beta^2/2)\overline{\Delta W^2} + \dots }
    \;=\;
    e^{-\beta\overline W + \Delta S/\KB}
    .
\label{507}
\end{gather}
The term $-\beta\overline W$ cancels out.  Keeping only leading
order in $\beta$, our high-temperature limit, and inserting Eq~(\ref{208}),
\begin{align}
    (\hbar\omega)^2
    (
    \overline{n}_\sssuparrow
    +
    \overline{n}_\sssdnarrow 
    )
    &\;=\;
     2(\KBT)^2\Delta S/\KB
     \,.
\label{205}
\end{align}
This set of $N$ two-state systems\cite{Reif} is a basic model in
introductory statistical mechanics, and  its  entropy is
\begin{align}
    S &\;=\;
    \KB\ln\frac{N!}{(N{-}{{n}})!\,{{n}}!}
\label{946}
\end{align}
when $n$ are in the upper state.  For $\Delta S$ under a change
$\overline{\Delta n} = \overline{n}_\sssuparrow {-} \overline{n}_\sssdnarrow$
we approximate
\begin{align}
    \Delta S
    &\;\approx\;
    \frac{\partial S}{\partial n}
    \,
    (\overline{n}_\sssuparrow
    {-}
    \overline{n}_\sssdnarrow)
    .
\label{576}
\end{align}
Approximating\cite{Reif} 
$\partial(\ln n!)/\partial n \approx \ln n$
to differentiate (\ref{946}), and using Boltzmann probabilities for the
states,
\begin{align}
    \frac{\partial S}{\partial {{n}}}
    &\;\approx\;
    \KB\ln\frac{N{-}{{n}}}{{{n}}}
    \;=\;
    \KB\ln(e^{\beta\hbar\omega})
    \;=\; \hbar\omega/T
    .
\label{842}
\end{align}
Inserting  this with (\ref{576}) into Eq~(\ref{205}) gives us
\begin{gather}
    \hbar\omega
    \,
    (
    \overline{n}_\sssuparrow
    +
    \overline{n}_\sssdnarrow 
    )
    \;=\;
     2\KBT 
     \,
    (\overline{n}_\sssuparrow
    {-}
    \overline{n}_\sssdnarrow)
    .
\label{498}
\end{gather}
Finally, 
the Boltzmann ratio of
subsystems in the upper state 
to those in the lower state 
is $n/(N{-}n) = \exp(-\hbar\omega/\KBT)$.
From this,
in our high-temperature limit where ${{n}}\approx N/2$,
i.e.~where $N/2 - {{n}}$ is small,
\begin{gather}
    \frac{\hbar\omega}{\KBT}
    \;=\;
    \ln\frac{N - {{n}}}{{{n}}}
    \;\approx\;
    (4/N)
    \Big(
    \frac N 2 - {{n}}
    \Big)
    .
\label{173}
\end{gather}
Following Einstein, the absorption and stimulated emission are
proportional to the light intensity and the number in the respective
states,
\begin{gather}
    \overline{n}_\sssuparrow
    =
    (N{-}{{n}}) BI
    ,\quad
    \overline{n}_\sssdnarrow
    =
    {{n}} \widetilde B I
\label{619}
\end{gather}
where $B$ is the coefficient of absorption and $\widetilde B$ is
Einstein's postulated coefficient of stimulated emission.  Combining
the last three equations we get
\begin{align}
    ( N{-}2{{n}})
    [(N{-}{{n}}) B
    +
    {{n}} \widetilde B
    ]
    &\;=\;
    N
    [
    (N{-}{{n}}) B
    -
    {{n}} \widetilde B
    ]
\label{836}
\end{align}
which is easily seen to be solved by
\begin{gather}
    \widetilde B
    \;=\;
    B
    .
\label{571}
\end{gather}
Stimulated emission is found to be necessary, and
furthermore we get Einstein's equality of the $B$ coeffients.

The quanta involved in stimulated emission of course must obey
Bose-Einstein statistics, so it is interesting that equivalence of
the $B$ coefficients as obtained here does not explicitly invoke
the Bose nature of the quanta, involving as it does only the work
on, and the free energy of, the collection of two-state systems.
By contrast, in Einstein's original argument\cite{Eisberg} Bose
statistics are introduced by the use of Planck's black-body spectrum,
and in second-quantized field theory\cite{ZimanAQM} stimulated
emission is a specific property of boson fields.

Historically, on the other hand, it has been emphasized by
Jaynes\cite{Reviewersuggest,Jaynes} and others that one can reproduce the rules of
absorption and stimulated emission by treating the light field
classically while retaining the quantum nature of the absorbers, as
we have done in a simple way here with our discrete subsystems.

\section{A derivation of Jarzynski's equality}
\label{Derivation}

Many derivations of the Jarzynski equality and related fluctuation
theorems by Crooks and others are available.  Here we rework a
particularly efficient quantum argument for the JE by 
D.~N.~Page,\cite{Page2012} which we follow with some brief commentary.
Suppose a system is described by a time-independent Hamiltonian
operator $H$.  In Gibbs equilibrium the probability of being in an
energy eigenstate ${\phi_i}$ with energy $E_i$ is
\begin{gather}
    p_i = Z(\beta)^{-1}e^{-\beta E_i} 
\label{007}
\end{gather}
where the partition function is
\begin{gather}
    Z(\beta) = \sum_i e^{ -\beta E_i}
     = e^{-\beta F}
\label{052}
\end{gather}
and $F$ is the Helmholtz free energy.  Now suppose an external
interaction causes $H$ to change over some time interval into a new
$H'$.  The $i$th eigenstate and energy change smoothly into some
new ${\phi'_i}$ and $E'_i$, but there will also be induced transitions.
Denote by $P_{i{\rightarrow}j}$ the probability that a system in
state ${\phi_i}$ is subsequently found in ${\phi'_j}$.  A specific
choice for microscopic work must be made\cite{Crooks2009} and our
quantum definition is $E'_j-E_i$, pairwise between states.  Then
\begin{align}
    \avg{ e^{ -\beta(E'-E)} }
    &=
    \sum_{i,j} 
    e^{-\beta(E'_j -E_i) }
    p_{i}
    P_{i{\rightarrow}j}
   \label{950}
 \\
    &=
    \sum_{i,j} 
    \big[e^{-\beta(E'_j -E_i) }\big]
    \big[
    e^{ -\beta E_i }
    /Z(\beta)
    \big]
    P_{i{\rightarrow}j}
 \\
    &=
    \sum_j 
    e^{-\beta E'_j}
    \big(
    \sum_i P_{i{\rightarrow}j}
    \big)
    /Z(\beta)
  \\
    &=
    Z'(\beta)/
    Z(\beta)
    .
\label{951}
\end{align}
We have used $\sum_i P_{i{\rightarrow}j}=1$.  With $F$ defined as
in (\ref{052}),
\begin{gather}
    \avg{ e^{ -\beta W} }
    \;=\;
    Z'(\beta)
    /
    Z(\beta)
    \;=\;
    e^{ -\beta(F'-F)}
\label{238}
\end{gather}
which is the desired equality.  The notation $Z'$ indicates use of
energies $E'_i$ in this partition function.

The above system is closed.  The fact that we keep $\beta$ unchanged
implies a very large heat capacity, as when our closed system
includes a heat reservoir.  However, if in (\ref{950}) we multiply each
$E'_j$ by $\beta'$ in the exponentials rather than $\beta$, Eq~(\ref{238}) 
easily generalizes to\cite{Page2012}
\begin{align}
    \avg{ e^{ -\beta'E'+\beta E} }
    &\;=\;
    Z'(\beta')/
    Z(\beta)
    \;=\;
    e^{ -\beta'F'+\beta F} 
    .
\label{952}
\end{align}
Now the final temperature of the disturbed and closed system can
be, and in general will be, different from the original.  The
generalized Eq~(\ref{952}) is strikingly symmetric in the initial and
final equilibria.  It is closely related to results such as the
Crooks fluctuation theorem\cite{Crooks1998,Crooks1999,Crooks2000,
EvansSearles2002}.

Eq~(\ref{952}) prompts some further comments.  The initial $\beta$ is
understandable, but how can the equality express the physics so
symmetrically in both $\beta$ and the eventual $\beta'$ that will
result infinitely far in the future?  In fact, our derivation of
the JE never committed to a direction of time.  Knowing the initial
$\beta$ and letting the system determine $\beta'$ at $t=-\infty$
is a matter of boundary conditions; we could instead choose a
$\beta'$ at $t=\infty$ and develop the system backwards through the
disturbance to find the correct $\beta$ at $t=-\infty$.  Time-reversed
trajectories also appear in classical discussions of the Jarzynski
and Crooks theorems.\cite{Crooks1998,Crooks2000}

The view of fluctuation and relaxation as time reversals of each
other has long been used\cite{Landau} in the context of linear
response and FDTs.  The JE and related theorems are exploiting such
a time-reversed view, but even for systems very far out of equilibrium.

\section{Conclusions}
\label{Conclusions}

We have considered two simple yet widely known results due to
Einstein, one classical and the other quantum. In each case 
we identified 
a well-characterized yet variable
amount of irreversible or nonequilibrium work $W$ is done on the
system initially in thermal equilibrium, so that Jarzynski's equality
applies.  Then, in each case, we expanded the JE average $\avg{
e^{-\beta W} }$ to second order in a cumulant expansion, and from
the resulting equation we were able to draw nontrivial conclusions.

We also adapted and presented D.~N.~Page's 
derivation and generalization of the JE, and took the opportunity for a
further discussion of the JE's basis in time reversal.  Time 
reversal also figured in a paradox (Section \ref{Macro limit}) 
in a macroscopic extrapolation of the average $\avg{ e^{-\beta W} }$.
The centrality of such issues has been articulated by Jarzynski 
and others.  Such paradoxes represent a 
warning to those who might hope to use the JE as freely as they would, 
say, the Gibbs distribution, and they show that the strength 
of these fluctuation theorems lies with microscopic systems.

On the other hand, the strategy for applying the JE that succeeded
here is reasonably straightforward, and seems promising as a model
for how Jarzynski's equality could be more widely useful in
physics.  Examples like the ones considered may help to establish
a place for the Jarzynski equality, and related theorems, in the
physicist's toolbox of useful principles.

\begin{acknowledgments}
I thank Md.~Kamrul Hoque Ome, a graduate student in one of my
recent classes, for bringing to my attention the Jarzynksi
derivation by D.~N.~Page.  I thank anonymous reviewers for several
useful suggestions, and my colleague J.~Thomas Dickinson for lively
input that prompted the discussion in Section \ref{Macro limit}.
I also thank Mark G.~Kuzyk, Michael McNeil Forbes, and Peter Engels
for many enjoyable discussions of issues in statistical and
quantum~physics that have been relevant to this work.
\end{acknowledgments}

%
%
%
%
%
%
%


\begin{thebibliography}{50}

\bibitem{Jarz1997a}
C.~Jarzynski, ``Nonequilibrium equality for free energy differences,''
Phys.~Rev.~Lett.~78, 2690-2693 (1997).

\bibitem{Jarz1997b}
C.~Jarzynski, ``Equilibrium free-energy differences from nonequilibrium
measurements: A master-equation approach,''
Phys.~Rev.~E 56, 5018-5035 (1997).


\bibitem{Crooks1998}
G.~E.~Crooks, ``Nonequilibrium measurements of free energy
differences for microscopically reversible markovian systems,''
J.~Stat.~Phys.~90, 1481-87 (1998). 


\bibitem{Crooks1999}
G.~E.~Crooks, ``Entropy production fluctuation theorem and the
nonequilibrium work relation for free energy differences,''
Phys.~Rev.~E 60, 2721-26 (1999).

\bibitem{Crooks2000}
G.~E.~Crooks, ``Path-ensemble averages in systems driven far from
equilibrium,'' Phys.~Rev.~E 61, 2361-66 (2000).


\bibitem{EvansSearles2002}
D.~J.~Evans and D.~J.~Searles, ``The fluctuation theorem,''
Adv.~Phys.~52, 1529-40 (2002).


\bibitem{Hummer2001}
G.~Hummer, A.~Szabo,
``Free energy reconstructon from nonequilibrium single-molecule
pulling experiments,'' Proc. Natl. Acad. Sci. U.S.A. 98 3658-61
(2001).

\bibitem{Bustamante2005}
C.~Bustamante, J.~Liphardt, F.~Ritort, ``The Nonequilibrium
Thermodynamics of Small Systems,'' Physics Today 58, 43-48 (2005).

\bibitem{Harris2007}
N.~C.~ Harris, Y.~ Song, C.-H.~ Kiang, ``Experimental free energy 
surface reconstruction from single-molecule force spectroscopy
using Jarzynski's equality,'' Phys.~Rev.~Lett.~99, 068101/1-4 (2007).

\bibitem{Jarz2011}
C.~Jarzynski, ``Equalities and inequalities: irreversibility and
the second law of thermodynamics at the nanoscale,'' 
Ann.~Rev.~Cond.~Matt.~Phys.~2, 329-351 (2011)


\bibitem{Peliti2011}
L.~Peliti, {\it Statistical Mechanics in a Nutshell}
(Princeton University Press, Princeton, 2011).


\bibitem{LuaGrosberg2005}
R,\ C.~Lua, A.~Y.~Grosberg, ``On practical applicability of the
Jarzynski relation in statistical mechanics: a pedagogical example'',
J. Phys. Chem. B 109, 6805-11 (2005).

\bibitem{Hijar2010}
H.~H\'ijar,  J.~M.~O.~de Z\'arate, ``Jarzynski's equality
illustrated by simple examples'', European J.~Phys.~31, 1097-1106 (2010).

\bibitem{Ribeiro2016}
W.~ L.~ Ribeiro, G.~ T.~ Landi, F.~ Semi\~ao, ``Quantum
thermodynamics and work fluctuations with applications to magnetic
resonance'', Am.~J.~Phys.~84, 948-957 (2016).


\bibitem{Landau}
L.~D.~Landau and E.~M.~Lifshitz, {\it Statistical Physics}, 2nd ed.
(Pergamon Press, Oxford, 1969).

\bibitem{Reif}
F.~Reif, 
{\it Fundamentals of statistical and thermal physics}
(McGraw-Hill, New York, 1965).

\bibitem{Marconi2008}
U.~M.~B.~Marconi, A.~Puglisi, L.~ Rondoni, A.~ Vulpiani,
``Fluctuation-dissipation: Response theory in statistical physics,''
Phys.~Reports 461, 111-196 (2008).

\bibitem{Chen2008a}
L.~Y.~Chen, ``On the Crooks fluctuation theorem and the Jarzynski
equality,'' J.~Chem.~Phys.~129, 091101/1-2 (2008).

\bibitem{Crooks2009}
G.~E.~Crooks, ``Comment regarding `On the Crooks fluctuation
theorem and the Jarzynski equality' and `Nonequilibrium
fluctuation-dissipation theorem of Brownian dynamics','' 
J.~Chem.~Phys.~129, 144113/1 (2008).

\bibitem {JarzRare2006}
C. Jarzynski, ``Rare events and the convergence of exponentially averaged
work values,'' Phys Rev E 73, 046105/1-10 (2006).

\bibitem{Eisberg}
R.~Eisberg and R.~Resnick, {\it Quantum physics of atoms, molecules,
solids, nuclei, and particles}, 2nd ed.  (Wiley, New York, 1984).

\bibitem {ZimanAQM}
J.~M.~Ziman, {\it Elements of Advanced Quantum Theory}
(Cambridge University Press, 1969).

\bibitem{Page2012}
D.~N.~Page, ``Generalized Jarzynski equality,'' arXiv:1207.3355v1 (2012).

\bibitem{Reviewersuggest}
I thank a reviewer for a useful suggestion regarding this point.

\bibitem {JTDobjection}
This point was raised in conversation by J.~Thomas Dickinson.

\bibitem{Jaynes}
C.~R.~Stroud and E.~T.~Jaynes, 
``Long-term solutions in semiclassical radiation theory,''
Phys.~Rev.~A 1, 106-121 (1970).

\bibitem{Nelson}
P.~Nelson, {\it Biological Physics: Energy, Information, Life},
(Freeman, New York, 2008).

\bibitem{JarzPers2007}
C. Jarzynski, personal communication (2007).

\bibitem{CrooksJarz2007}
G.~E.~Crooks, C.~Jarzynski, ``Work distribution for the adiabatic
compression of a dilute and interacting classical gas,''
Phys Rev E 75, 021116/1-4 (2007).

\end{thebibliography}
\end{document}